\documentclass[twocolumn, prd, aps, showpacs, preprintnumbers]{revtex4-2}

\usepackage{amsmath}
\usepackage[table]{xcolor}
\newcommand{\SU}{\text{SU}}
\newcommand{\U}{\text{U}}
\usepackage{tikz}
\usetikzlibrary{"arrows", "automata", "backgrounds", "calendar", "chains", "matrix", "mindmap", "patterns", "petri", "shadows", "shapes.geometric", "shapes.misc", "spy", "trees"}
\usetikzlibrary{arrows,shapes}
\usetikzlibrary{decorations.pathmorphing} 
\usetikzlibrary{trees}
\usetikzlibrary{matrix,arrows} 	
\usetikzlibrary{positioning}	
\usetikzlibrary{calc,through}
\usepackage{amssymb}
\newcommand{\grayentry}[1]{\colorbox{gray!25}{$\displaystyle #1$}}
\usetikzlibrary{decorations.pathreplacing} 
\usepackage{pgffor}	
\usepackage{comment}
\definecolor{blau}{RGB}{0,51,160}
\definecolor{taronja}{RGB}{255,108,12}
\usetikzlibrary{decorations.pathmorphing}
\usepackage[bookmarksnumbered=true]{hyperref} 
\hypersetup{
    colorlinks = true,
    linkcolor = blau,
    anchorcolor = blue,
    citecolor = blau,
    filecolor = blau,
    urlcolor = blau
    }
\usetikzlibrary{decorations.markings}
\tikzset{
    vector/.style={decorate, decoration={snake}, draw},
    scalarloop/.style={dashed,draw=black, postaction={decorate},
        decoration={markings,mark=at position .75 with {\arrow[draw=black]{<}}}},
    scalarloop1/.style={dashed,draw=black, postaction={decorate},
        decoration={markings,mark=at position .25 with {\arrow[draw=black]{<}}}},
	provector/.style={decorate, decoration={snake,amplitude=2.5pt}, draw},
	antivector/.style={decorate, decoration={snake,amplitude=-2.5pt}, draw},
    fermion/.style={draw=black, postaction={decorate},
        decoration={markings,mark=at position .55 with {\arrow[draw=black]{>}}}},
    fermioncyan/.style={draw=black, postaction={decorate},
        decoration={markings,mark=at position .55 with {\arrow[draw=cyan]{<}}}},
    fermiondif/.style={draw=black, postaction={decorate},
        decoration={markings,mark=at position .7 with {\arrow[draw=black]{>}}}},
            fermiondif2/.style={draw=black, postaction={decorate},
        decoration={markings,mark=at position .7 with {\arrow[draw=black]{<}}}},
    fermionend/.style={draw=black, postaction={decorate},
        decoration={markings,mark=at position 1 with {\arrow[draw=black]{>}}}},
    fermionuchannel2/.style={draw=black, postaction={decorate},
        decoration={markings,mark=at position .4 with {\arrow[draw=black]{>}}}},
    scalardif/.style={dashed,draw=black, postaction={decorate},
        decoration={markings,mark=at position .7 with {\arrow[draw=black]{>}}}},
    scalarend/.style={dashed,draw=black, postaction={decorate},
        decoration={markings,mark=at position 1 with {\arrow[draw=black]{>}}}},
    fermionbar/.style={draw=black, postaction={decorate},
        decoration={markings,mark=at position .55 with {\arrow[draw=black]{<}}}},
    fermionnoarrow/.style={draw=black},
    gluon/.style={decorate, draw=black,
        decoration={coil,amplitude=4pt, segment length=5pt}},
    scalar/.style={dashed,draw=black, postaction={decorate},
        decoration={markings,mark=at position .55 with {\arrow[draw=black]{>}}}},
    scalarcyan/.style={dashed,draw=black, postaction={decorate},
        decoration={markings,mark=at position .55 with {\arrow[draw=cyan]{>}}}},
    scalaruchannel1/.style={dashed,draw=black, postaction={decorate},
        decoration={markings,mark=at position .7 with {\arrow[draw=black]{>}}}},
                  scalaruchannel2/.style={dashed,draw=black, postaction={decorate},
        decoration={markings,mark=at position .4 with {\arrow[draw=black]{>}}}},
    scalarbar/.style={dashed,draw=black, postaction={decorate},
        decoration={markings,mark=at position .55 with {\arrow[draw=black]{<}}}},
    scalarnoarrow/.style={dashed,draw=black},
    electron/.style={draw=black, postaction={decorate},
        decoration={markings,mark=at position .55 with {\arrow[draw=black]{>}}}},
	bigvector/.style={decorate, decoration={snake,amplitude=4pt}, draw},
}
\usetikzlibrary{arrows.meta}

\usepackage{flushend}

\tikzstyle{block} = [draw, rectangle, 
    minimum height=3em, minimum width=6em]
\tikzset{
    cross/.pic = {
    \draw[rotate = 45] (-#1,0) -- (#1,0);
    \draw[rotate = 45] (0,-#1) -- (0, #1);
    }
}
\tikzstyle{block} = [draw, rectangle, 
    minimum height=3em, minimum width=6em]

    \usepackage{xparse}

\usepackage{amsmath, amssymb}
\usepackage{hyperref}
\usepackage{flushend} 
    
\NewDocumentCommand\semiloop{O{black}mmmO{}O{above}}
{%
\draw[#1] let \p1 = ($(#3)-(#2)$) in (#3) arc (#4:({#4+180}):({0.5*veclen(\x1,\y1)})node[midway, #6] {#5};)
}
\usepackage{balance}

\begin{document}

\title{\Large{Universal seesaw mechanism for fermion masses and strong CP problem}}

\author{Pavel Fileviez P\'erez$^{1}$, Clara Murgui$^{2}$}
\email{pxf112@case.edu,clara.murgui@cern.ch}
\affiliation{
$^{1}$Physics Department and Center for Education and Research in Cosmology and Astrophysics,
Case Western Reserve University, Cleveland, OH 44106, USA \\
$^{2}$Theoretical Physics Department, CERN,
1 Esplanade des Particules, CH-1211 Geneva 23, Switzerland
}

\preprint{CERN-TH-2026-147}

\begin{abstract}
We propose a framework for fermion mass generation based on a universal seesaw. The Standard Model is extended by an Abelian gauge symmetry acting on right-handed fermions, together with vector-like fermions and scalar fields. The ordinary Yukawa couplings are forbidden, except for the top-quark coupling to the Higgs, which is allowed at the renormalizable level and remains unsuppressed. The charged-fermion hierarchies then arise from mixing with the vector-like sector, and light neutrino masses emerge from a neutral sector Majorana seesaw. CP is exact in the ultraviolet and broken spontaneously by scalar vacuum expectation values. The resulting CP-violating phase is transmitted to the quark sector and generates the CKM phase, while a Nelson-Barr structure, realized through the universal seesaw block form, keeps the physical QCD vacuum angle zero at tree level. Consistency with EDM bounds beyond tree level requires moderately suppressed Yukawa couplings between SM doublets to the vector-like sector. If the leading higher-dimensional operators are unsuppressed, the same requirement can favor a low breaking scale for the new Abelian symmetry. In this regime the vector-like fermions can lie at the TeV scale, with suppressed mixing with the electroweak sector.
 This framework provides a simple setting in which the hierarchies of charged-fermion masses, neutrino masses, and CP-violating parameters can be accounted for within a common extension of the Standard Model.

\end{abstract}

\maketitle

\section{Introduction}
%
The origin of fermion masses and flavor hierarchies remains one of the central open questions in particle physics. 
In the Standard Model (SM), charged fermion masses arise from Yukawa interactions with the Higgs field, while neutrino masses require physics beyond the minimal SM. The observed spectrum spans many orders of magnitude, from sub-eV neutrino masses~\cite{Fukuda:1998mi,Ahmad:2002jz,Planck:2018vyg,KATRIN:2022ith} to the electroweak-scale top quark~\cite{ParticleDataGroup:2024cfk}. Since the corresponding Yukawa couplings are free parameters, the SM offers no explanation for this pattern, usually referred to as the flavor puzzle. At the same time, the strong CP phase, which is not expected a priori to be small~\cite{Baluni:1978rf,Crewther:1979pi}, is severely constrained by the non-observation of the neutron electric dipole moment (EDM)~\cite{Abel:2020pzs,Pospelov:1999mv,Liang:2023jfj}.

Vector-like fermions provide a well-motivated extension of the SM in which the flavor puzzle may find a natural interpretation. Since their masses are not tied to electroweak symmetry breaking, they can lie above the electroweak scale and evade direct searches, while still leaving indirect imprints through their mixing with ordinary fermions. This logic is familiar from Froggatt-Nielsen-type constructions~\cite{Froggatt:1978nt}, where hierarchical effective Yukawa couplings arise after integrating out heavy messengers charged under a broken flavor symmetry. Universal seesaw models~\cite{Berezhiani:1985in,Chang:1986bp,Rajpoot:1987fca,Davidson:1987mh} provide a concrete vector-like fermion realization of this idea, in which light fermion masses arise only after integrating out the heavy vector-like states and are therefore suppressed by the seesaw block structure. Variants of this mechanism have been explored in a variety of gauge and flavor settings~\cite{Babu:1989rb,Davidson:1993xn,Babu:1999me,Jana:2021tlx,Panuluh:2024pau,Babu:2025xgi}. Recent work has revived this perspective in TeV-scale constructions based on vector-like fermion sectors protected by suitable flavor symmetries~\cite{Arkani-Hamed:2026wwy,Greljo:2025mwj}.

In this Letter, we propose a framework that provides a common origin for fermion hierarchies and the suppression of strong CP violation. The theory extends the SM by a chiral Abelian gauge symmetry acting on the right-handed SM fermions. This symmetry forbids the ordinary SM Yukawa couplings, except for the top-quark Yukawa, while allowing the SM fermions to mix with heavier states that are vector-like under the SM gauge group. After spontaneous symmetry breaking, the charged fermion spectrum is generated through a universal seesaw mechanism, controlled by moderately suppressed Yukawa couplings between the SM electroweak doublets and the vector-like sector. The top quark is singled out by a direct Higgs coupling that accounts for its large mass. Neutrinos can be Majorana particles and acquire naturally smaller masses through the corresponding neutral sector seesaw. The model should not be viewed as a complete predictive theory of flavor: in the absence of an additional horizontal symmetry, the detailed CKM matrix and the precise intergenerational mass pattern depend on the flavor orientation of the Yukawa and vector-like mass matrices. Rather, the model explains the order-of-magnitude structure of the observed mass hierarchies. 

The flavor puzzle and the strong CP problem are intertwined at the level of the quark mass matrices: the same structures responsible for an order-one Cabibbo-Kobayashi-Maskawa (CKM) phase can also feed CP violation into the physical QCD vacuum angle, $\bar\theta_{\rm QCD}$, unless they are suitably protected. It is therefore natural for a theory of flavor to address, or at least control, the origin of strong CP violation; see, for example, Refs.~\cite{Calibbi:2016hwq,Davidi:2018sii,delaVega:2021ugs,Cherchiglia:2024ssz,Babu:2025xgi} for related approaches. A central feature of our construction is that CP is exact in the ultraviolet and broken only spontaneously by scalar vacuum expectation values. The resulting relative phase is transmitted to the light quark sector and can generate a CKM phase. At the same time, a distinctive Nelson-Barr-type mechanism~\cite{Nelson:1983zb,Nelson:1984hg, Barr:1984qx}, realized through the universal-seesaw block structure rather than through complex corrections to an otherwise allowed SM Yukawa matrix, ensures that the physical strong CP phase vanishes at tree level and can remain sufficiently small after radiative corrections. In this way, the charge assignment under the new Abelian gauge symmetry enforces a universal seesaw structure that relates several otherwise unexplained hierarchies: the charged-fermion spectrum, the light neutrino scale, the exceptional top mass, and the smallness of $\bar\theta_{\rm QCD}$.

\section{Theoretical Framework}
\label{sec:theory}
%
The model is based on the gauge symmetry
\begin{equation}
{\cal G}_R = \SU(3)_C \otimes \SU(2)_L \otimes  \U(1)_Y \otimes \U(1)_R.
\end{equation}
We first describe the charge assignments for one SM generation. The left-handed SM fermions are neutral under the new Abelian symmetry,
\begin{equation}\label{eq:LHSM}
\begin{split}
    q_L &\sim (3,2,1/6,0),  \quad
    \ell_L \sim (1,2,-1/2,0),
\end{split}
\end{equation}
where the entries denote quantum numbers under ${\cal G}_R$. The right-handed fermions transform as
\begin{equation}
\label{eq:rcharges}
\begin{split}
u_R &\sim (3,1,2/3,r_U),\quad 
d_R \sim (3,1,-1/3,-r), \\
e_R &\sim (1,1,-1,-r),  \quad
\nu_R \sim (1,1,0,r).    
\end{split}
\end{equation}
For $r_U=r$, the SM fermion sector is anomaly free for arbitrary $r$. We refer to the new symmetry as $\U(1)_R$ because it acts non-trivially on the right-handed SM fermions.
The Higgs doublet is assigned charge
\begin{equation}
H \sim (1,2,1/2,r_H).
\end{equation}
The ordinary SM Yukawa interactions require $r_H=r=r_U$. We instead take $r_H\neq r$, thereby forbidding these couplings. 
To generate fermion masses, we introduce new fermions that are vector-like under the SM gauge group but chiral under $\U(1)_R$,
\begin{equation}
\begin{aligned}
U_R &\sim (3,1,2/3,-R_U), \quad & U_L &\sim (3,1,2/3,L_U), \\
D_R &\sim (3,1,-1/3,R),  \quad  & D_L &\sim (3,1,-1/3,-L), \\
E_R &\sim (1,1,-1,R),  \quad  & E_L &\sim (1,1,-1,-L), \\
N_R &\sim (1,1,0,-R),  \quad  & N_L &\sim (1,1,0,L).
\end{aligned}
\end{equation}
This sector is anomaly free for arbitrary $L=L_U$ and $R=R_U$. We keep this general choice for now. The new fermions acquire large masses after the spontaneous breaking of $\U(1)_R$ by the vacuum expectation value (VEV) of the scalar field 
\begin{equation}\label{eq:S}
S \sim (1,1,0,L+R),
\end{equation}
via the following interactions, 
\begin{equation}
\begin{split}
-{\cal L} &\supset \left(\lambda_U \bar U_L U_R +\lambda_N \bar N_L N_R \right) S
\\
&\quad + \left( \lambda_D \bar{D}_L  D_R + \lambda_E \bar{E}_L  E_R  \right) S^* + \text{h.c.},
\end{split}
\end{equation}
where $L\neq -R$ is assumed. The bridge interactions connecting the SM and vector-like sectors are fixed by choosing
\begin{equation}
    r_H = -R \neq r, \quad \text{and}\quad L=r.
\end{equation}
The first condition allows
\begin{equation}
\begin{split}
-{\cal L} &\supset  \ y_u  \bar{q}_L \tilde{H} U_R + y_d  \bar{q}_L H D_R  \\
& \quad + y_\nu \bar{\ell}_L \tilde{H} N_R + y_e  \bar{\ell}_L H E_R + \text{h.c.},
\end{split}
\end{equation}
with $\tilde H = i \sigma_2 H^*$, while the second allows
\begin{equation}
\begin{split}
-{\cal L}  &\supset   m_U \bar{U}_L u_R   + m_D \bar{D}_L d_R  \\
&\quad + m_N \bar{N}_L \nu_R   + m_E \bar{E}_L e_R   +  \text{h.c.}.
\end{split}
\end{equation}
Since $r_H \neq r$, the SM Yukawa couplings remain forbidden, and the light fermions acquire masses through a universal seesaw mechanism. 

The extension to three generations is straightforward. For example, in the charged-lepton sector,
\begin{equation}
\begin{split}
  \!\!\!\!  -{\cal L} &\supset \mathsf{Y}_e^{ij} \bar \ell_L^i H E_R^j  + \mathsf{m}_E^{ij} \bar{E}_L^i e_R^j + \lambda_E^{ij} \bar{E}_L^i E_R^j S^{*} + \text{h.c.},
\end{split}
\end{equation}
with $\mathsf{Y}_e$, $\mathsf{m}_E$, and $\lambda_E$ promoted to $3\times3$ matrices. In the following sections we specify the additional charge assignments and flavor structures needed to obtain realistic fermion masses and to implement the Nelson-Barr solution to the strong CP problem.
\section{Fermion Masses}
\label{sec:masses}
For each charged-fermion sector, the mass matrix ($6\times 6$) can be written in the basis
$(f_L,F_L)$ and $(f_R,F_R)$ as
\begin{equation}
    {\cal M}_f \equiv \begin{pmatrix}
\mathsf{M}_{11}^f & \mathsf M_{12}^f\\
\mathsf M_{21}^f & \mathsf M_{22}^f
\end{pmatrix} =
    \frac{1}{\sqrt{2}}
    \begin{pmatrix}
    0_{3\times 3} & \mathsf{Y}_f\, v_H \\
    \sqrt{2}\, \mathsf{m}_F & \mathsf{\lambda}_F\, v_S
    \end{pmatrix},
\end{equation}
where $f$ denotes the SM fermions and $F$ their vector-like (under the SM gauge group) partners. 
We parametrize the bridge mass as $\mathsf{m}_F = \lambda_f \, \Lambda_m$, thereby factoring out the common mass scale $\Lambda_m$ and defining $\lambda_f$ as the corresponding dimensionless Yukawa matrix.
If $\lambda_F$ is invertible, the light fermion masses are generated by the universal seesaw relation
\begin{equation}\label{eq:seesaw}
\mathsf{m}_f^{\rm eff}
= - \mathsf{M}_{12}^f (\mathsf{M}_{22}^f)^{-1} \mathsf{M}_{21}^f = -
\mathsf{Y}_f
\mathsf{\lambda}_F^{-1}
\mathsf{\lambda}_f \frac{v_H \Lambda_m}{v_S}.
\end{equation}
Thus the scale of the charged-fermion masses is controlled by the ratio
$v_H\Lambda_m /v_S$,
up to products of dimensionless matrices. Although $\mathsf{M}_{21}^f\propto \Lambda_m$ originates from a bare mass term in the Lagrangian, its smallness relative to $\mathsf{M}_{22}^f\propto v_S$ is technically natural: in the limit $\mathsf{M}_{21}^f\to0$, the fermion sector acquires an enhanced flavor symmetry, which protects this hierarchy against large radiative corrections.

For the first two generations $(\ell=1,2)$, we take common $\U(1)_R$ charges, $r_{U \ell}= r_\ell$, $R_\ell=R_{U\ell}$ and $L_\ell = L_{U\ell}$. With this choice, the matrices $\mathsf{Y}_f$, $\lambda_f$, and $\lambda_F$ entering Eq.~\eqref{eq:seesaw} are generic in the corresponding two-dimensional flavor space. The non-linear structure of the seesaw relation can generate hierarchies of several orders of magnitude around the common seesaw scale. For instance, for entries in the range $0.1-1$, first and second generation masses of charged fermions can be accommodated for $v_H\Lambda_m / v_S={\cal O}(0.1)$ GeV.

While the seesaw scale $v_H\Lambda_m/v_S$ can account for the first two charged-fermion generations, it is too small to generate the top-quark mass with perturbative couplings, motivating a different charge assignment for the third generation.

We choose the third-generation charges such that the ordinary up-type Yukawa coupling to $u_R^3$ is allowed, while the down-quark and charged-lepton sectors retain the universal seesaw form. The explicit fermion mass textures implied by the charge assignment in Table~\ref{tab:charges} are given in the Supplemental Material. In the up sector, the effective light mass matrix is then
\begin{equation}
\mathsf m_u^{\rm eff}
=
\mathsf M_{11}^u
-
\mathsf M_{12}^u
(\mathsf M_{22}^u)^{-1}
\mathsf M_{21}^u ,
\label{eq:upmass}
\end{equation}
where \(\mathsf M_{11}^u\) is rank one and gives the dominant contribution to the top-quark mass. Denoting the direct Higgs coupling by $\widetilde{\mathsf Y}_u^{i3} \bar Q_L^i \tilde H u_R^3$, one finds
\begin{equation}
m_t \simeq \frac{v_H}{\sqrt2} \sqrt{|\widetilde{\mathsf Y}_u^{13}|^2 + |\widetilde{\mathsf Y}_u^{23}|^2 + |\widetilde{\mathsf Y}_u^{33}|^2},
\end{equation}
while the remaining up-type masses arise from the seesaw contribution in Eq.~\eqref{eq:upmass}. 
In addition, the charge assignment allows entries proportional to the $\U(1)_R$ breaking scale in $\mathsf{M}_{21}^u$, which can reduce the seesaw suppression of one additional eigenvalue and help accommodate the charm mass. 
By contrast, in the down-quark and charged-lepton sectors, $\mathsf M_{11}^{d,e}=\mathsf{0}_{3\times 3}$. Since the remaining mass blocks are rank three, their three nonzero masses arise entirely through the seesaw structure and are correspondingly suppressed.

The neutrino sector can be treated analogously. If lepton number is conserved, neutrinos are Dirac fermions and their masses lie near the charged-fermion seesaw scale, which is unacceptably large for democratic Yukawa couplings. We instead allow lepton-number violation, adopting the charge assignment listed in Table~\ref{tab:charges}, so that the neutral fermions realize a Majorana seesaw~\cite{Minkowski:1977sc,Gell-Mann:1979vob,Mohapatra:1979ia,Yanagida:1979as}. For a single generation, in the basis $(\nu_L,\nu_R^c,N_L,N_R^c)^T$, 
\begin{equation}
{\cal M}_\nu = 
\begin{pmatrix}
0 & \mathsf{v}^T\\
\mathsf{v} & \mathsf M_H
\end{pmatrix},
\label{eq:Mmaj}
\end{equation}
where $\mathsf{M}_H$ is a $3\times 3$ matrix containing the heavy neutral-fermion Dirac and Majorana masses (see Supplemental Material), and
\begin{equation}
  \mathsf{v}^T = \frac{v_H}{\sqrt{2}} (0, Y_L ,Y_\nu),
\end{equation}
contains the electroweak insertions.
Thus, while charged fermions obtain masses through a Dirac universal seesaw, see Eq.~\eqref{eq:seesaw}, linear with the light entries $\mathsf{Y}_f v_H$ and suppressed by $\Lambda_m / v_S$, 
neutrinos acquire masses through a Majorana seesaw,
\begin{equation}
m_\nu^{\rm light}
\simeq
-
\mathsf{v}^T \mathsf{M}_H^{-1}\mathsf{v} ={\cal O} ( Y^2 v_H^2 / v_S),
\label{eq:nu-seesaw}
\end{equation}
quadratic in the electroweak insertion $Y v_H$ and controlled by the inverse heavy neutral-sector Majorana scale. In the last estimate we have taken the entries of
$\mathsf M_H$ to be of order $v_S$, and $Y_\nu$ and $Y_L$ of order $Y$. For $Y ={\cal O}(1)$, Eq.~\eqref{eq:nu-seesaw} requires $v_S$ near the canonical seesaw scale. However, because of the quadratic dependence on $Y$, suppressed neutrino Yukawa couplings allow a much lower Majorana scale.

Thus, the charge assignment in Table~\ref{tab:charges} enables a unified framework that accommodates the charged-fermion hierarchies, the exceptional top-quark mass, and neutrino masses suppressed by a Majorana seesaw.

\begin{table}[t]
    \centering
    \renewcommand{\arraystretch}{1.2}
    \begin{tabular}{|c |c |c || c |c |c|}
    \hline
         \cellcolor{gray!10}Fields & $\ell=1,2$ & 3
        &  \cellcolor{gray!10}Fields & $\ell=1,2$ & 3 \\
        \hline
        \cellcolor{gray!10}$q_L$ & 0 & 0 &
        \cellcolor{gray!10}$\ell_L$ & 0 & 0 \\

        \cellcolor{gray!10}$u_R$ & $1$ & \ -$1$ \ &
        \cellcolor{gray!10}$d_R$ & -$1$  & \ $3$ \ \\

        \cellcolor{gray!10}$e_R$ & -$1$ & $3$ &
        \cellcolor{gray!10}$\nu_R$ & $1$ & -$3$ \\

        \cellcolor{gray!10}$U_L$ & $1$ & \ -$3$ \ &
        \cellcolor{gray!10}$U_R$ & -$1$ & \ -$3$ \ \\

        \cellcolor{gray!10}$D_L$ & -$1$ & $3$ &
        \cellcolor{gray!10}$D_R$ & $1$ & $1$ \\
        
        \cellcolor{gray!10}$E_L$ & -$1$& $3$ & \cellcolor{gray!10}$E_R$ & $1$& $1$\\

        \cellcolor{gray!10}$N_L$ & $1$& -$3$& 
        \cellcolor{gray!10}$N_R$ &  -$1$& -$1$\\
        \hline
    \end{tabular}
    \caption{$\U(1)_R$ charge assignments for the fermion fields. The scalar charges are $H\sim(1,2,1/2,-1)$ and $S\sim(1,1,0,2)$. The column $\ell=1,2$ corresponds to the first two generations, and the column labelled $3$ to the third generation.}
    \label{tab:charges}
\end{table}
%
\section{A Nelson-Barr-type mechanism}
\label{sec:NB}
%
In a generic theory of quark masses, complex Yukawa matrices contribute to the physical QCD vacuum angle as follows
\begin{equation}\label{eq:strongCP}
\bar\theta_{\rm QCD} = \theta_{\rm QCD}  + \arg \{ \det {\cal M}_u \} + \arg \{ \det  {\cal M}_d \},
\end{equation}
where $\theta_{\rm QCD}$ denotes the topological contribution multiplying
$G\tilde G$. The non-observation of the neutron electric dipole moment requires $|\bar\theta_{\rm QCD}| \lesssim 10^{-10}$~\cite{Abel:2020pzs,Pospelov:1999mv,Liang:2023jfj}. We now show that the universal seesaw structure described above admits a Nelson-Barr-type realization~\cite{Nelson:1983zb,Barr:1984qx,Nelson:1984hg}, in which $\bar \theta_{\rm QCD} = 0$ at tree level while a CKM phase can be consistently generated.

We assume that CP is an exact symmetry of the ultraviolet. Then $\theta_{\rm QCD}=0$, and all  couplings are real before spontaneous symmetry breaking. A single scalar $S$ is not sufficient to generate a physical CP phase, since its phase is {\it eaten} by the massive $\U(1)_R$ gauge boson. Two scalar fields with the same quantum numbers, however, allow for a gauge-invariant relative phase. We therefore take
\begin{equation}
S,S' \sim (1,1,0,L_\ell+R_\ell).
\end{equation}
The relevant phase-dependent terms in the scalar potential are
\begin{equation}
    V \supset - \mu^2 \, S^* S' + \lambda (S^*  S')^2 + \text{h.c.},
\end{equation}
with real $\mu^2$ and $\lambda$. The parameter $\mu^2$ should be understood as an effective mass-squared parameter multiplying ($S^* S'+\text{h.c.})$, which includes not only the bare bilinear contribution but also all additional phase-independent radial contributions that generate the same relative phase dependence ({\it e.g.} $(S^*S)S^*S'$). Writing
\begin{equation}
\langle S \rangle
=
\frac{v_{S}}{\sqrt{2}} \text{e}^{\text{i}\theta_{S}},
\quad
\langle S' \rangle
=
\frac{v_{S}'}{\sqrt{2}} \text{e}^{\text{i}\theta_{S}'},
\end{equation}
the minimization conditions fix the relative phase,
\begin{equation}
\cos(\theta_{S}-\theta_{S}')
=
\frac{\mu^2}{2\lambda v_{S}v_{S}'} ,
\end{equation}
so that a nontrivial CP-violating vacuum is obtained for suitable parameters, {\it i.e.} $\theta_{S}-\theta_{S}'\neq 0,\pi$.

In the down-quark and charged-lepton sectors, the resulting complex phase enters in the heavy vector-like block
 \begin{equation}
 \mathsf{M}_{22}^f =\tfrac{1}{\sqrt{2}}\big( \mathsf{\lambda}_F \, v_S\,  \text{e}^{\text{i} \theta_S} + \mathsf{\lambda}_F' \, v_{S'}\, \text{e}^{\text{i} \theta_{S'}}\big),
 \end{equation}
rendering it generically complex. However, because the universal seesaw structure has a vanishing upper-left block, $\mathsf{M}_{11}^f=0$, the determinant of the full mass matrix is independent of the entries in $\mathsf{M}_{22}^f$. Therefore,
\begin{equation}
    \arg \{ \det {\cal M}_d \}=0,
\end{equation}
at tree level.
The up-quark sector contains a direct rank-one SM-like block responsible for the top mass. Nevertheless, with the charge assignment in Table~\ref{tab:charges}, the complex entries, located in the third column of $\mathsf{M}_{21}^u$ and the first two columns of $\mathsf{M}_{22}^u$, still drop out of the determinant. Expanding along the vector-like top column and then along the first two columns of ${\cal M}_u$ gives 
\begin{equation}\label{eq:detMu}
\begin{split}
 \det  {\cal M}_u
 &=\frac{v_H^3 \Lambda^2_m}{\sqrt{2}^3}  \mathsf{m}_U^{33} \det \! \begin{pmatrix} \lambda_u^{11} & \lambda_u^{12} \\ \lambda_u^{21} & \lambda_u^{22} \end{pmatrix} \\
 &\qquad \qquad \quad \times \det  \begin{pmatrix} \mathsf{\tilde Y}_u^{13} & \mathsf{Y}_u^{11} & \mathsf{Y}_u^{12} \\
    \mathsf{\tilde Y}_u^{23} & \mathsf{Y}_u^{21} & \mathsf{Y}_u^{22} \\
    \mathsf{\tilde Y}_u^{33} & \mathsf{Y}_u^{31} & \mathsf{Y}_u^{32} \end{pmatrix}.
\end{split}
\end{equation}
All entries in this expression are real and, consequently,
\begin{equation}\label{eq:treezero}
    \arg \{ \det {\cal M}_u \} =0.
\end{equation}
Therefore, $\bar \theta_{\rm QCD} = 0$ at tree level. 

The same structure nevertheless generates a CKM phase. The complex vector-like mass matrix $\mathsf{M}_{22}^f$
 enters the light fermion masses through the universal seesaw relation. 
We note that, unlike in ordinary Nelson-Barr models, generating an order-one CKM phase does not require the bridge mass scale $[\mathsf m_f]$ to be comparable to the vector-like mass scale $[\mathsf M_{22}^f]$, since here the CKM phase is controlled by the flavor-nontrivial complex structure of $\mathsf{M}_{22}^f$. Here and in what follows, square brackets denote the typical order of magnitude of the entries of a matrix.  

Having established the tree-level Nelson-Barr-type mechanism, we now turn to its quality. Radiative corrections and higher-dimensional operators can perturb the Nelson-Barr texture and induce a nonzero $\bar \theta_{\rm QCD}$. We denote by $\delta\mathsf{M}_f$ a small correction to the tree-level fermion mass matrix $\mathsf{M}_f$,
\begin{equation}
\delta \mathsf{M}_f \equiv
\begin{pmatrix}
\delta \mathsf{M}_{11}^f & \delta \mathsf{M}_{12}^f\\
\delta \mathsf{M}_{21}^f & \delta \mathsf{M}_{22}^f
\end{pmatrix}.
\end{equation}
Expanding to first order in $\delta\mathsf{M}_d$, the induced shift from the down-quark sector is
\begin{equation}\label{eq:corrections}
\begin{split}
\Delta \bar \theta_{\rm QCD}^{(d)} &=
{\rm Im}\big \{ {\rm Tr}\{
- (\mathsf{M}_{21}^d)^{-1}\mathsf{M}_{22}^d(\mathsf{M}_{12}^d)^{-1}
\,\delta \mathsf{M}_{11}^d  \\
&\hspace{1 cm}
+ (\mathsf{M}_{21}^d)^{-1}\delta \mathsf{M}_{21}^d
+ (\mathsf{M}_{12}^d)^{-1}\delta \mathsf{M}_{12}^d
\}\big\} .
\end{split}
\end{equation}
Among these terms, the most dangerous correction is usually a nonzero upper-left block, $\delta\mathsf{M}_{11}^d$, corresponding to a direct SM-like Yukawa interaction, since its contribution is enhanced by the inverse effective light fermion mass matrix, see Eq.~\eqref{eq:seesaw}. The up-quark sector can be treated analogously, although the analytic expressions are less transparent because the direct top Yukawa coupling gives a nonzero rank-one contribution to $\mathsf M_{11}^u$.
The qualitative lesson, however, remains the same: complex  corrections to the light block must be suppressed in order not to regenerate an unacceptably large $\bar\theta_{\rm QCD}$.

As discussed in the Supplemental Material, loop-induced complex masses can remain compatible with the stringent EDM bound provided the Yukawa couplings $\mathsf{Y}_f$, which connect the SM doublets to the vector-like sector, are moderately suppressed. We refer to $\mathsf{Y}_f$ as the electroweak-mixing Yukawa couplings. This observation motivates the benchmark regime considered below. 

\begin{figure}[t]
    \centering
\includegraphics[width=0.99\columnwidth]{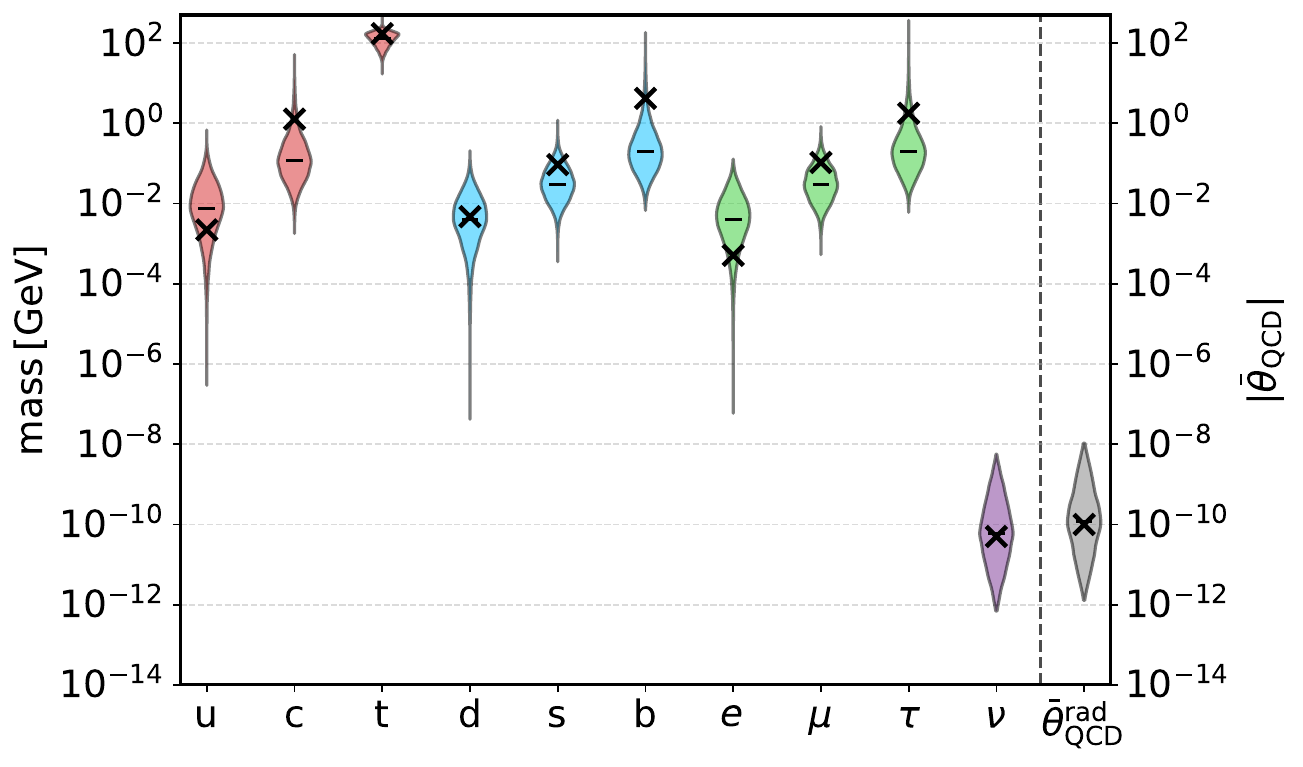}
    \caption{
    Monte Carlo illustration of the hierarchy-generating mechanism within the model presented in this Letter. The violin
    distributions are obtained from $10^4$ randomly generated points for which the relevant mass matrices are nonsingular. They show the singular values of the effective charged fermion
    mass matrices, see Eqs.~\eqref{eq:seesaw} and~\eqref{eq:upmass}, the light neutrino mass estimate, see Eq.~\eqref{eq:nu-seesaw}, and the radiatively induced contribution
    $|\bar \theta_{\rm QCD}^{\rm rad}|$, see the Supplemental Material. The nonzero texture entries are sampled
   logarithmically within the order-of-magnitude ranges described in the text. Crosses
    indicate reference values: the observed charged-fermion masses, a neutrino scale
    $m_\nu\simeq0.05\,{\rm eV}$, and
    $|\bar \theta_{\rm QCD}|=10^{-10}$. The plot should therefore be interpreted
    as an illustration of hierarchy generation, not as a statistical flavor fit.}
    \label{fig:hierarchy-scan}
\end{figure}

Taking $[\mathsf{Y}_f] = {\cal O}(10^{-3})$, order-one values for the remaining Yukawa couplings, and a mild seesaw ratio $\Lambda_m/v_S\sim0.1$, the charged-fermion seesaw scale is
\begin{equation}
[\mathsf{Y}_f] v_H \frac{\Lambda_m}{v_S} = {\cal O}(10)\text{ MeV}.
\end{equation}
The nonlinear matrix structure in
$\lambda_F^{-1}\lambda_f$ can enhance or suppress individual eigenvalues
around this scale, while the top quark is generated by its direct Higgs
coupling. The same moderately small $[\mathsf Y_f]$ improves the Nelson-Barr quality by suppressing the radiatively generated scalar portal.  It also lowers the light neutrino masses through the quadratic dependence $m_\nu\sim Y^2 v_H^2/v_S$, allowing
sub-eV neutrino masses for a low-intermediate Majorana scale. Such moderately suppressed electroweak-mixing Yukawa couplings can be interpreted as controlled breaking of an approximate chiral flavor symmetry acting on the SM electroweak doublets, in the spirit of recent vector-like flavor constructions~\cite{Arkani-Hamed:2026wwy}. 
The discussion above can be summarized by a simple Monte Carlo illustration.
We sample logarithmically the nonzero entries allowed by the textures, taking $[\mathsf Y_f] = [10^{-3},10^{-2}]$, $[\lambda_f],[\lambda_F]= [0.1,1]$, $\Lambda_m/v_S = [10^{-2},10^{-1}]$, $v_S = [10^{7}, 10^{9}]$ GeV, and impose no CKM optimization. The purpose of the scan is therefore not to fit flavor, but to test the hierarchy-generating structure of the model. The resulting distributions, shown in Fig.~\ref{fig:hierarchy-scan}, illustrate that the allowed textures naturally populate the observed charged-fermion, neutrino, and $\bar\theta_{\rm QCD}$ scales. The GeV-scale seesaw-generated fermions, especially the charm, bottom, and tau, lie toward the upper tails of the distributions, indicating the mild eigenvalue hierarchies or flavor alignment required for the heaviest states not directly coupled to the Higgs.

Higher dimensional operators, starting at dimension-five, can also induce complex corrections to the quark mass matrices. If such operators are suppressed or sufficiently aligned, the intermediate-scale benchmark illustrated in Fig.~\ref{fig:hierarchy-scan} remains viable. If, instead, they are present with unsuppressed Wilson coefficients, requiring their contribution to $\bar\theta_{\rm QCD}$ to remain acceptably small favors a low $\U(1)_R$-breaking scale, corresponding to vector-like fermions with masses in the ${\cal O}(1\!-\!10) \text{ TeV}$ range (see the Supplemental Material). Such states would be phenomenologically interesting, since they could be accessible through direct production at future high-energy colliders. Nevertheless, the suppressed mixing between the SM electroweak doublets and the vector-like sector in this benchmark makes electroweak precision tests challenging. We note, however, that a low spontaneous-CP-breaking scale can be cosmologically nontrivial, since the associated domain walls must either be inflated away or avoided through a modified thermal history, such as high-temperature symmetry non-restoration~\cite{McNamara:2022lrw,Dvali:1995cc,Dvali:1996zr}.

\section{SUMMARY}
\label{sec:conclusions}
In this Letter, we presented a framework for fermion mass generation based on a universal seesaw mechanism, in which the Standard Model fermions obtain masses through mixing with heavy vector-like states. A chiral Abelian gauge symmetry acting on the right-handed sector forbids the usual SM Yukawa couplings, with the exception of the top Yukawa, and organizes the seesaw structure. The first two generations of charged fermions are naturally suppressed by the universal seesaw, while the top quark is distinguished by an unsuppressed Higgs coupling. In the neutral sector, lepton-number violation leads to a Majorana seesaw, allowing neutrino masses to be separated from the charged-fermion Dirac-seesaw scale.

A key feature of this framework is that it admits a Nelson-Barr-type realization of spontaneous CP violation. CP is exact in the ultraviolet and broken only by scalar vacuum expectation values. The resulting phase is transmitted to the light quark sector through the inverse vector-like mass matrix, generating a CKM phase while keeping $\bar\theta_{\rm QCD}=0$ at tree level. Radiative corrections and higher-dimensional operators then provide important probes of the quality of the mechanism. Requiring these effects to remain compatible with EDM bounds motivates moderately suppressed couplings between the SM electroweak doublets and the new vector-like sector. If the leading higher-dimensional operators are unsuppressed, the same requirement can favor a relatively low scale for the breaking of the new Abelian symmetry, implying vector-like fermions in the ${\cal O}(1\!-\!10) \text{ TeV}$ range. Such states could be directly accessible at future high-energy colliders, while their suppressed mixing with the SM electroweak doublets makes precision electroweak tests challenging.
The model is best viewed not as a complete theory of flavor, but as a theory of hierarchies: within a simple Abelian extension of the Standard Model, it provides a common origin for the charged-fermion spectrum, the neutrino scale, the possibility of CP violation in the quark and lepton sectors, and the smallness of $\bar\theta_{\rm QCD}$.

{\textit{Acknowledgments:} We dedicate this article to our mentor and collaborator Prof. Mark B. Wise.}

\onecolumngrid

\appendix

\setcounter{section}{0}
\setcounter{equation}{0}
\setcounter{figure}{0}
\setcounter{table}{0}

\renewcommand{\thesection}{S\arabic{section}}
\renewcommand{\theequation}{S\arabic{section}-\arabic{equation}}
\renewcommand{\thefigure}{S\arabic{figure}}
\renewcommand{\thetable}{S\arabic{table}}

\vspace{1.0cm}

\begin{center}
{\large{\bf{Supplemental Material}}}
\end{center}
\setcounter{equation}{0}
\section{Gauge Charges and Mass Textures}
In this Supplemental Material we give the charge assignments used in the main text and derive the corresponding mass-matrix textures. The gauge symmetry is
\begin{equation}
{\cal G}_R =
\SU(3)_C \otimes \SU(2)_L \otimes \U(1)_Y \otimes \U(1)_R .
\end{equation}
For the first and second generations, denoted by \(\ell=1,2\), we take
\begin{equation}
r_{U\ell}=r_\ell,\qquad
L_{U\ell}=L_\ell,\qquad
R_{U\ell}=R_\ell .
\end{equation}
For the third generation we allow \(r_3\neq r_{U3}\). Anomaly cancellation is preserved if
\begin{equation}
R_{U3} = -r_3,\qquad
L_{U3} = L_3,\qquad
r_{U3} = -R_3.
\end{equation}
We further choose
\begin{equation}
r_3=L_3,
\end{equation}
and relate the third-generation charges to those of the first two generations by
\begin{equation}
R_3=R_\ell,\qquad
r_3=-r_\ell-2R_\ell .
\end{equation}
The resulting charge assignment is summarized in Table~\ref{tab:chargesgen}, where \(x\equiv R_\ell/r_\ell\).

\begin{table}[h]
    \centering
    \renewcommand{\arraystretch}{1.2}
    \begin{tabular}{|c |c |c || c |c |c|}
    \hline
         \cellcolor{gray!10}Fields & $\ell=1,2$ & 3
        &  \cellcolor{gray!10}Fields & $\ell=1,2$ & 3 \\
        \hline
        \cellcolor{gray!10}$q_L$ & 0 & 0 &
        \cellcolor{gray!10}$\ell_L$ & 0 & 0 \\

        \cellcolor{gray!10}$u_R$ & $1$ & \ $-x$ \ &
        \cellcolor{gray!10}$d_R$ & $-1$  & \ $1+2x$ \ \\

        \cellcolor{gray!10}$e_R$ & $-1$ & $1+2x$ &
        \cellcolor{gray!10}$\nu_R$ & $1$ & $-1-2x$ \\

        \cellcolor{gray!10}$U_L$ & $1$ & \ $-1-2x$ \ &
        \cellcolor{gray!10}$U_R$ & $-x$ & \ $-1-2x$ \ \\

        \cellcolor{gray!10}$D_L$ & $-1$ & $1+2x$ &
        \cellcolor{gray!10}$D_R$ & $x$ & $x$ \\
        
        \cellcolor{gray!10}$E_L$ & $-1$& $1+2x$ & \cellcolor{gray!10}$E_R$ & $x$& $x$\\

        \cellcolor{gray!10}$N_L$ & $1$& $-1-2x$& 
        \cellcolor{gray!10}$N_R$ &  $-x$& $-x$\\
        \hline
    \end{tabular}
    \caption{$\U(1)_R$ charge assignments for the fermion fields, with $x\equiv R_\ell/r_\ell$. The scalar charges are $H \sim (1,2,1/2,-x)$ and $S \sim (1,1,0,1+x)$, in units of $r_\ell$. For generic $x$, the neutral sector may conserve lepton number. For $x=1$, lepton-number-violating terms are allowed and Majorana neutrino masses arise.}
    \label{tab:chargesgen}
\end{table}
\subsection{Down-type quarks and charged leptons}
In the down quark sector, one has
\begin{equation}\label{eq:downsector}
    \begin{split}
     \hspace{-0.4cm}  -{\cal L} &\supset \mathsf{Y}_d^{ij} \bar q_L^i H D_R^j + \mathsf{m}_D^{\ell \ell'} \bar D_L^{\ell} d_R^{\ell'}+\mathsf{m}_D^{33} \bar D_L^3  d_R^3   + \lambda_D^{\ell i} \bar D_L^{\ell} S^*  D_R^{i} + \lambda_D^{3 i} \bar D_L^{3} S D_R^i + \text{h.c.},
    \end{split}
\end{equation}
where $\ell,\ell' =1,2$ label the first and second generations, while $i,j=1,2,3$. An analogous structure holds in the charged lepton sector. 
The corresponding textures are
\begin{equation}
\mathsf M_{11}^{d,e}=0_{3\times3},
\end{equation}
and
\begin{equation}
\mathsf M_{12}^{d,e}
=
\frac{v_H}{\sqrt2}
\begin{pmatrix}
\mathsf Y_{d,e}^{11} & \mathsf Y_{d,e}^{12} & \mathsf Y_{d,e}^{13}\\
\mathsf Y_{d,e}^{21} & \mathsf Y_{d,e}^{22} & \mathsf Y_{d,e}^{23}\\
\mathsf Y_{d,e}^{31} & \mathsf Y_{d,e}^{32} & \mathsf Y_{d,e}^{33}
\end{pmatrix},
\quad 
\mathsf M_{21}^{d,e}
=
\Lambda_m
\begin{pmatrix}
\lambda_{d,e}^{11} & \lambda_{d,e}^{12} & 0\\
\lambda_{d,e}^{21} & \lambda_{d,e}^{22} & 0\\
0 & 0 & \lambda_{d,e}^{33}
\end{pmatrix},
\quad
\mathsf M_{22}^{d,e}
=
\frac{v_S}{\sqrt2}
\begin{pmatrix}
\lambda_{D,E}^{11} & \lambda_{D,E}^{12} & \lambda_{D,E}^{13}\\
\lambda_{D,E}^{21} & \lambda_{D,E}^{22} & \lambda_{D,E}^{23}\\
\lambda_{D,E}^{31} & \lambda_{D,E}^{32} & \lambda_{D,E}^{33}
\end{pmatrix}.
\end{equation}
The universal-seesaw relation discussed in the main Letter implies that the rank of the light mass matrix satisfies the bound 
\begin{equation}\label{eq:rank}
    \text{rank}(\mathsf{m}_{d,e})
    \leq
    \min \! \big\{
    \text{rank}(\mathsf{M}^{d,e}_{12}),
    \text{rank}(\mathsf{M}^{d,e}_{21}),
    \text{rank}(\mathsf{M}^{d,e}_{22})
    \big\}.
\end{equation}
The matrices $\mathsf{M}_{12}^{d,e}$, $\mathsf{M}_{21}^{d,e}$ and $\mathsf{M}_{22}^{d,e}$ are rank three, and therefore there is enough freedom to generate three light fermion masses spanning about three orders of magnitude around the universal-seesaw scale in the main Letter.

For the down-quark sector, the vanishing upper-left block implies
\begin{equation}
\det{\cal M}_d
=
-\det(\mathsf M_{12}^d\,\mathsf M_{21}^d).
\end{equation}
Since $\mathsf{M}_{12}^d$ and $\mathsf{M}_{21}^d$ are real in the CP basis,
\begin{equation}
\arg \{ \det{\cal M}_d \}=0 .
\end{equation}
The same texture applies to charged leptons, although they do not contribute to $\bar\theta_{\rm QCD}$.
\subsection{Up-type quarks}
In the up quark sector, the allowed interactions are instead
\begin{equation}\label{eq:upsector}
\begin{split}
     \hspace{-0.4cm} -{\cal L} &\supset \mathsf{Y}_u^{i\ell}\bar q_L^i \tilde H U_R^{\ell} + \mathsf{\tilde Y}_u^{i3} \bar q_L^i \tilde H u_R^3  + \mathsf{m}_U^{\ell \ell'}\bar U_L^{\ell} u_R^{\ell'}  + \lambda_U^{\ell \ell'} \bar U_L^\ell S U_R^{\ell'} \\
     &\qquad + \lambda_U^{3\ell} \bar U_L^3 S^* U_R^{\ell} +  \mathsf{m}_U^{33} \bar U_L^3 U_R^3 + \xi_u^{\ell 3} \bar U_L^\ell S u_R^3  +  \xi_u^{33} \bar U_L^3 S^* u_R^3  + \text{h.c.}.
\end{split}
\end{equation}
We note that $\mathsf{\tilde Y}_u$ corresponds to ordinary SM-like Yukawa couplings for the top quark, and $\xi_u$ to the allowed {\it bridge} interactions with $S$, which are absent in the down-quark and charged-lepton sector.

The textures are then given by
\begin{equation}
\begin{alignedat}{4}
\mathsf M_{11}^u
&=
\frac{v_H}{\sqrt2}
\begin{pmatrix}
0 & 0 & \widetilde{\mathsf Y}_u^{13} \\
0 & 0 & \widetilde{\mathsf Y}_u^{23} \\
0 & 0 & \widetilde{\mathsf Y}_u^{33}
\end{pmatrix},
\qquad&
\mathsf M_{12}^u
&=
\frac{v_H}{\sqrt2}
\begin{pmatrix}
\mathsf Y_u^{11} & \mathsf Y_u^{12} & 0 \\
\mathsf Y_u^{21} & \mathsf Y_u^{22} & 0 \\
\mathsf Y_u^{31} & \mathsf Y_u^{32} & 0
\end{pmatrix},
\\[0.8em]
\mathsf M_{21}^u
&=
\Lambda_m
\begin{pmatrix}
\lambda_u^{11} & \lambda_u^{12} &
\dfrac{\xi_u^{13}}{\sqrt2}\dfrac{v_S}{\Lambda_m} \\
\lambda_u^{21} & \lambda_u^{22} &
\dfrac{\xi_u^{23}}{\sqrt2}\dfrac{v_S}{\Lambda_m} \\
0 & 0 &
\dfrac{\xi_u^{33}}{\sqrt2}\dfrac{v_S}{\Lambda_m}
\end{pmatrix},
\qquad&
\mathsf M_{22}^u
&=
\frac{v_S}{\sqrt2}
\begin{pmatrix}
\lambda_U^{11} & \lambda_U^{12} & 0 \\
\lambda_U^{21} & \lambda_U^{22} & 0 \\
\lambda_U^{31} & \lambda_U^{32} &
\sqrt2\,\dfrac{\mathsf m_U^{33}}{v_S}
\end{pmatrix}.
\end{alignedat}
\end{equation}
Although $\mathsf{M}_{12}^u$ has rank two, the presence of the nonzero block $\mathsf{M}_{11}^u$ lifts the pure seesaw structure, and the rank bound in Eq.~\eqref{eq:rank} no longer applies. The effective mass matrix for the light up type quarks is instead
\begin{equation}
\mathsf m_u^{\rm eff}
=
\mathsf M_{11}^u
-
\mathsf M_{12}^u
(\mathsf M_{22}^u)^{-1}
\mathsf M_{21}^u .
\end{equation}
The first term has rank one and gives the dominant third-generation up-type mass, whereas the second term generates the seesaw masses for the lighter generations. Hence, the top quark mass corresponds to the largest singular value, given by
\begin{equation}
m_t \simeq \frac{v_H}{\sqrt2} \sqrt{(\widetilde{\mathsf Y}_u^{13})^2 + (\widetilde{\mathsf Y}_u^{23})^2 + (\widetilde{\mathsf Y}_u^{33})^2} .
\end{equation}
Since $\Lambda_m\leq v_S$, Yukawa couplings of comparable size can make one vector-like fermion lighter than the others. This reduces the seesaw suppression for one eigenvalue, allowing the charm quark mass to be larger than the masses of its second-generation charged-fermion partners.

The above textures lead to the following determinant of the full up-quark mass matrix
\begin{equation}\label{eq:detMuS}
   \det {\cal M}_u
  = \frac{v_H^3 \Lambda_m^2 \mathsf{m}_U^{33}}{\sqrt{2}^3}  \det \! \begin{pmatrix} 0 & 0 & \mathsf{\tilde Y}_u^{13}  & \mathsf{Y}_u^{11} & \mathsf{Y}_u^{12} & 0 \\  0 & 0 & \mathsf{\tilde Y}_u^{23}  & \mathsf{Y}_u^{21} & \mathsf{Y}_u^{22} & 0 \\ 0 & 0 & \mathsf{\tilde Y}_u^{33}  & \mathsf{Y}_u^{31}  & \mathsf{Y}_u^{32}  & 0 \\
    \lambda_{u}^{11}  & \lambda_{u}^{12}  & \grayentry{\tilde \lambda_u^{13}} & \grayentry{\tilde \lambda_{U}^{11}} &  \grayentry{\tilde \lambda_{U}^{12}} & 0 \\ \lambda_{u}^{21}  & \lambda_{u}^{22}  & \grayentry{\tilde \lambda_u^{23}} &   \grayentry{\tilde \lambda_{U}^{21}} & \grayentry{\tilde \lambda_{U}^{22}} & 0 \\ 0 & 0 & \grayentry{\tilde \lambda_u^{33}} &  \grayentry{\tilde \lambda_{U}^{31}} & \grayentry{\tilde \lambda_{U}^{32}} & 1\end{pmatrix},
\end{equation}
where
\begin{equation}
\begin{split}
   &\tilde \lambda_{U}^{\ell \ell'} \equiv \frac{1}{\sqrt{2}\Lambda_m} \bigg( v_S \, \lambda_{U}^{\ell \ell'}\text{e}^{\text{i}\theta_{S}}  + v_{S'}\,  \lambda_{U}'^{\ell \ell'} \text{e}^{\text{i}\theta_{S'}} \bigg),\\
    &\tilde \lambda_u^{\ell 3} \equiv \frac{1}{\sqrt{2}\Lambda_m} \bigg( v_S\, \xi_u^{\ell 3} \text{e}^{\text{i}\theta_{S}} + v_S'\, {\xi'_u}^{\ell 3} \text{e}^{\text{i}\theta_{S'}} \bigg), \\
    &\tilde \lambda_u^{33} \equiv \frac{1}{\sqrt{2}\, \Lambda_m}\bigg(v_S \xi_u^{33} \text{e}^{-\text{i}\theta_{S}} + v_{S'} {\xi'_u}^{33}  \text{e}^{-\text{i}\theta_{S'}} \bigg),
\end{split}
\end{equation}
with unprimed and primed Yukawa couplings parametrizing the couplings to the two scalar fields $S$ and $S'$, respectively. These quantities are generically complex once both $S$ and $S'$ acquire CP-violating vacuum expectation values. Hence, the shaded entries in Eq.~\eqref{eq:detMuS} are complex after spontaneous CP breaking. Nevertheless, they do not enter the determinant. Expanding first along the last column and then along the first two columns of the remaining determinant yields the real, factorized determinant given in the main Letter. Only real entries appear in that expression, and therefore
\begin{equation}
\arg \{ \det{\cal M}_u \}=0 .
\end{equation}
Together with the down-sector result above, this implies that the flavor contribution to \(\bar\theta_{\rm QCD}\) vanishes at tree level.
\subsection{Neutrinos}
For a generic $x$ in Table~\ref{tab:chargesgen}, the neutral sector may conserve lepton number, yielding three seesaw-suppressed Dirac neutrinos together with three heavy Dirac neutral leptons. For order-one Yukawa couplings, the Dirac neutrino masses would be suppressed by the same universal seesaw scale as the charged fermions and would typically be too large. One possibility is to introduce additional hierarchies in the neutral bridge masses, $[\mathsf m_N]\ll [\mathsf m_{f^+}]$, where $f^+$ represents the charged fermions. 

A more appealing option is to allow lepton-number violation, so that neutrinos are Majorana particles.
This occurs for the charge assignment $x=1$, equivalently
\begin{equation}
R_{12}=r_{12}.
\end{equation}
In this case, lepton-number-violating neutral-sector terms are allowed and $\U(1)_R$ charges in Table~\ref{tab:chargesgen} simplify to the charge assignment employed in the main Letter.
Under this charge assignment, the neutrino sector is described by the following Dirac terms:
\begin{equation}
    - \mathcal{L}_\nu \supset
\mathsf{Y}_\nu^{ij}\,\bar{\ell}^{i}_L \tilde H N^j_R
+\mathsf{m}_N^{\ell \ell^{'}} \bar{N}_L^\ell \nu_R^{\ell^{'}} +   \mathsf{m}_N^{33} \bar{N}_L^3 \nu_R^3 + \lambda_N^{\ell i}\,\bar{N}^\ell_L\, S\, N^{i}_R  + \lambda_N^{3 i} \, \bar N^3_L \, S^* N^i_R +\mathrm{h.c.},
\label{eq:Lnu}
\end{equation}
as well as the following Majorana mass terms:
\begin{equation}
\begin{split}
-{\cal L}_\nu &\supset \mathsf{Y}_{L}^{i\ell} \ell_L^T C \text{i}\sigma_2 H N_L^\ell + 
\tfrac{1}{2}\lambda_{LL}^{\ell\ell'} (N_L^\ell)^T C N_L^{\ell'} S^*
+ \lambda_{LL}^{\ell3} (N_L^\ell)^T C N_L^3 S
+ \tfrac{1}{2}\lambda_{RR}^{ij}(N_R^i)^T C N_R^j S
\\
&\qquad
+ \mathsf m_{Rr}^{i\ell}(N_R^i)^T C\nu_R^\ell
+\tfrac{1}{2} \lambda_{rr}^{\ell\ell'}(\nu_R^\ell)^T C\nu_R^{\ell'} S^*
+ \lambda_{rr}^{\ell3}(\nu_R^\ell)^T C\nu_R^3 S
+ \text{h.c.}.
\end{split}
\end{equation}
The Majorana bilinears are symmetric under interchange of the two fermion fields, and the corresponding coupling matrices should be understood symmetrically in flavor space.
The complete neutral lepton mass matrix in the
basis $(\nu_L,\,\nu^c_R,\,N_L,\,N^c_R)^T$ for a single generation (e.g. a light generation) is then given, in block form, by
\begin{equation}
    {\cal M}_\nu = \begin{pmatrix} 
    0 & \mathsf{v}^T \\ \mathsf{v} & \mathsf{M}_H
    \end{pmatrix},
\label{eq:MmajS}
\end{equation}
where 
\begin{equation}
    \mathsf{M}_H = \begin{pmatrix} \lambda_{rr} \frac{v_S}{\sqrt{2}} & m_N &  m_{Rr} \\ m_N & \lambda_{LL} \frac{v_S}{\sqrt{2}} & \lambda_N \frac{v_S}{\sqrt{2}} \\
    m_{Rr} & \lambda_N \frac{v_S}{\sqrt{2}} & \lambda_{RR} \frac{v_S}{\sqrt{2}} \end{pmatrix} \!\! , \qquad  \mathsf{v} = \frac{v_H}{\sqrt{2}}\begin{pmatrix} 0 \\ Y_L \\  Y_\nu \end{pmatrix}.
\end{equation}
If $\mathsf{M}_H$ is invertible, since $|\mathsf{v}| \ll |\mathsf{M}_H|$, then
\begin{equation}
m_\nu^{\rm light}
\simeq
-\mathsf v^T\mathsf M_H^{-1}\mathsf v
=
- \frac{v_H^2}{2} \left( Y_L^2 (\mathsf{M}_H^{-1})_{22} + 2 Y_L Y_\nu(\mathsf{M}_H^{-1})_{23} + Y_\nu^2 (\mathsf{M}_H^{-1})_{33}\right).
\end{equation}
For heavy-sector entries of order $v_S$, and $Y_\nu \sim Y_L \sim Y$, this gives
$m_\nu^{\rm light}
=
{\cal O}(Y^2 v_H^2 / v_S)$, up to order-one factors.
\setcounter{equation}{0}
\section{The Quality of the Nelson-Barr Mechanism}
Radiative corrections and higher-dimensional operators can perturb the Nelson-Barr texture and spoil the vanishing tree-level prediction for $\bar\theta_{\rm QCD}$. 
In the universal seesaw limit, the most dangerous corrections are those that fill the light block $\mathsf M_{11}^q$, since they are weighted by the inverse of the seesaw-generated light fermion mass matrix,
\begin{equation}
\Delta \bar\theta_{\rm QCD}^{(q)}
\supset
\text{Im} \big \{ \text{Tr} \{ (\mathsf{m}_q^{\rm eff})^{-1}\delta \mathsf{M}_{11}^q
\} \big \}.
\end{equation}
Hence, even small corrections to $\mathsf M_{11}^q$ can generate an unacceptably large contribution to $\bar\theta_{\rm QCD}$ unless they are sufficiently suppressed or aligned with the effective light quark mass matrix.

A representative one-loop contribution to $\delta \mathsf{M}_{11}^q$ is shown in Fig.~\ref{fig:loop}. It involves the scalar portal interactions
\begin{equation}
V\supset \lambda_4^{ij} H^\dagger H S_i^\dagger S_j ,
\qquad \text{with }S_i=\{S,S'\}.
\end{equation}
Parametrically, this correction gives
\begin{equation}\label{eq:thetarad}
\Delta \bar \theta_{\rm QCD}
\sim
{\cal O} \left(\frac{\lambda_4}{16\pi^2}\right),
\end{equation}
where $\lambda_4$ denotes a representative portal coupling. Above we took into account that, with two independent CP-breaking spurions, the alignment between  the loop-induced $\delta\mathsf{M}_{11}^q$ and the tree-level seesaw mass matrix $\mathsf{m}_q^{\rm eff}$ is not generic. For order-one Yukawa couplings, the neutron EDM bound requires roughly
\begin{equation}
\lambda_4 \lesssim 10^{-8},
\end{equation}
which can be in tension with the portal coupling radiatively generated by the same fermion interactions~\cite{Perez:2023zin},
\begin{equation}
\delta\lambda_4
\sim
\frac{N_c}{16\pi^2}
{\rm Tr}
\{
\mathsf Y_f^\dagger \mathsf Y_f
\lambda_F^\dagger\lambda_F
\}.
\end{equation}
This tension is relaxed for moderately small electroweak-mixing Yukawa couplings, $\mathsf Y_f$, for which this conservative estimate can naturally be compatible with the present EDM bound.

\begin{figure}[t]
\begin{equation*}
\begin{gathered}
\begin{tikzpicture}[line width=1.5 pt,node distance=1 cm and 1 cm]
\coordinate[label=left:$q_L$](dR);
\coordinate[right = 1 cm of dR](v1);
\coordinate[right = 2 cm of v1](v2);
\coordinate[right= 1 cm of v2](dL);
\coordinate[right= 0.5 cm of v2,label=below:$D_L$](dLlab);
\coordinate[right= 1 cm of v1](vaux);
\coordinate[above = 1 cm of vaux](v4);
\coordinate[above left = 1 cm of v4,label=left:$\langle H \rangle$](Xi);
\coordinate[above right = 1 cm of v4,label=right:$\langle S \rangle$](Xj);
\coordinate[right = 1 cm of v1,label=below:$D_R$](DLlabel);
\coordinate[above = 0.75 cm of v1,label=left:$H$](Xlabel);
\coordinate[above = 0.75 cm of v2,label=right:$S$](Xlabel);
\coordinate[right = 1 cm of v2](v5);
\coordinate[right = 1 cm of v5,label=right:$d_R$](dRR);
\draw[scalarnoarrow] (v4)--(Xi);
\draw[scalarnoarrow] (v4)--(Xj);
\draw[fermion](dR)--(v1);
\draw[fermion](v1)--(v2);
\draw[fermion](v2)--(dL);
\draw[fermion](v5)--(dRR);
\draw[fill=black] (v1) circle (.05cm);
\draw[fill=black] (v2) circle (.05cm);
\draw[fill=black] (v4) circle (.05cm);
\draw[fill=black] (v5) circle (.05cm);
\semiloop[scalarloop]{v1}{v2}{0};
\semiloop[scalarloop1]{v1}{v2}{0};
\end{tikzpicture}
\end{gathered}
\end{equation*}
\caption{One-loop diagram contributing to the light block $\mathsf{M}_{11}^d$.}
\label{fig:loop}
\end{figure}

Higher-dimensional operators provide an additional source of quality violation. In particular, the leading effects can arise from corrections to the $\mathsf{M}_{11}^q$ block generated by dimension-five operators of the form
\begin{equation}
\frac{c_{i \ell}}{\Lambda}\bar q_L^i H d_R^{\ell} S
+
\frac{c'_{i \ell}}{\Lambda}\bar q_L^i H d_R^{\ell} S'
+\text{h.c.} .
\end{equation}
After spontaneous CP breaking, the simultaneous presence of both operators generally induces a complex correction to the light quark mass matrix proportional to the physical relative phase between the two singlet vevs, $\delta\equiv \theta_S-\theta_{S'}$. Requiring $\Delta\bar\theta_{\rm QCD}
\lesssim 10^{-10}
$
then gives the parametric bound
\begin{equation}
v_S \lesssim
100~{\rm TeV}
\left(\frac{\Lambda}{10^{19}~{\rm GeV}}\right)
\left(\frac{10^{-5}}{m_d/m_t}\right)
\left(\frac{0.1}{c \sin\delta}\right),
\end{equation}
where $c$ denotes a representative Wilson coefficient. Thus, if these operators are present with unsuppressed Wilson coefficients, the $\U(1)_R$-breaking scale (and therefore the scale of spontaneous CP violation) is pushed toward comparatively low values. Such a low scale would require correspondingly smaller Yukawa couplings $\mathsf{Y}_\nu$ and $\mathsf{Y}_L$ in the neutrino sector in order to reproduce the observed neutrino masses. 

\bibliography{refs}


\end{document}